\documentclass[manuscript]{acmart}

\definecolor{darkulred}{RGB}{150,0,0}
\definecolor{charcoal}{HTML}{333333}
\definecolor{brick}{HTML}{8B3A3A}
\definecolor{warmmist}{HTML}{F5F5F0}
\definecolor{pearlgray}{HTML}{EAEAEA}
\definecolor{palesage}{HTML}{E6E8D3}

\definecolor{darkspringgreen}{rgb}{0.09, 0.45, 0.27}
\definecolor{cyan(process)}{rgb}{0.0, 0.72, 0.92}
\definecolor{darkpastelblue}{rgb}{0.0, 0.72, 0.92}

\AtBeginDocument{%
  }

\renewcommand\footnotetextcopyrightpermission[1]{}
\acmISBN{978-1-4503-XXXX-X/2018/06}

\usepackage{multirow}
\usepackage{enumitem}
\begin{document}

\title{How Relevance Emerges: A Mechanistic Analysis of LoRA Fine-Tuning in Reranking LLMs}

\author{Atharva Nijasure}
\author{Tanya Chowdhury}
\author{James Allan}
\email{{anijasure,tchowdhury,allan}@cs.umass.edu}
\affiliation{%
  \institution{\\Center for Intelligent Information Retrieval, Manning College of Information and Computer Sciences, University of Massachusetts Amherst}
  \city{Amherst}
  \state{MA}
  \country{USA}
}

\renewcommand{\shortauthors}{Nijasure et al.}

\begin{abstract}
We conduct a behavioral exploration of LoRA fine-tuned LLMs for Passage Reranking to understand how relevance signals are learned and deployed by Large Language Models. By fine-tuning Mistral-7B, LLaMA3.1-8B, and Pythia-6.9B on MS MARCO under diverse LoRA configurations, we investigate how relevance modeling evolves across checkpoints, the impact of LoRA rank (1, 2, 8, 32), and the relative importance of updated MHA vs. MLP components. Our ablations reveal which layers and projections within LoRA transformations are most critical for reranking accuracy. These findings offer fresh explanations into LoRA’s adaptation mechanisms, setting the stage for deeper mechanistic studies in Information Retrieval. All models used in this study can be found here\footnote{\url{https://huggingface.co/AtharvaNijasureUMass/}}.
\end{abstract}
\maketitle

% \vspace{-3mm}
\section{Background}
\vspace{-1.5mm}

\begin{description}[left=0pt]
    \item[\textbf{Motivation}:] Fine-tuning Large Language Models (LLMs) using Low-Rank Adaptation (LoRA) has become a popular approach for adapting pre-trained transformers to Information Retrieval tasks. However, the inner workings of these fine-tuned models remain largely opaque, limiting our ability to understand how and where relevance signals are encoded and used. In this study, we attempt to demistify the workings of the LoRA updates that grant ranking ability to LLMs. This work contributes toward our long-term goal of uncovering novel latent features encoded in the MLP layers of LLMs and integrating them into conventional statistical ranking models.
     \item[\textbf{Reranking Task}:]
    We use the Tevatron repository\footnote{\url{https://github.com/texttron/tevatron}} to fine-tune LLMs for passage reranking on MS MARCO~\cite{nguyen2016ms}, similar to the existing work~\cite{ma2024fine,chowdhury2024probing}.
    Given a query-document pair, the model learns to predict a relevance score via a cross-entropy loss, using hard negatives for contrastive learning. We evaluate performance using nDCG@10 on the TREC DL19 and DL20 benchmarks~\cite{craswell2020overview}.
    \item[\textbf{LoRA Fine-Tuning}:]
    Instead of updating all model parameters, LoRA~\cite{hu2022lora} injects a lightweight, low-rank approximation into existing layers, drastically reducing computational overhead and storage needs. 
    This low-rank module is merged with the base model at inference time, often achieving parity with full fine-tuning in tasks such as passage reranking~\cite{shuttleworth2024lora,ma2023fine}.
\end{description}

% \vspace{-3mm}
\section{Experiments}
\begin{figure}[h]
    \centering
    \includegraphics[width=0.5\linewidth]{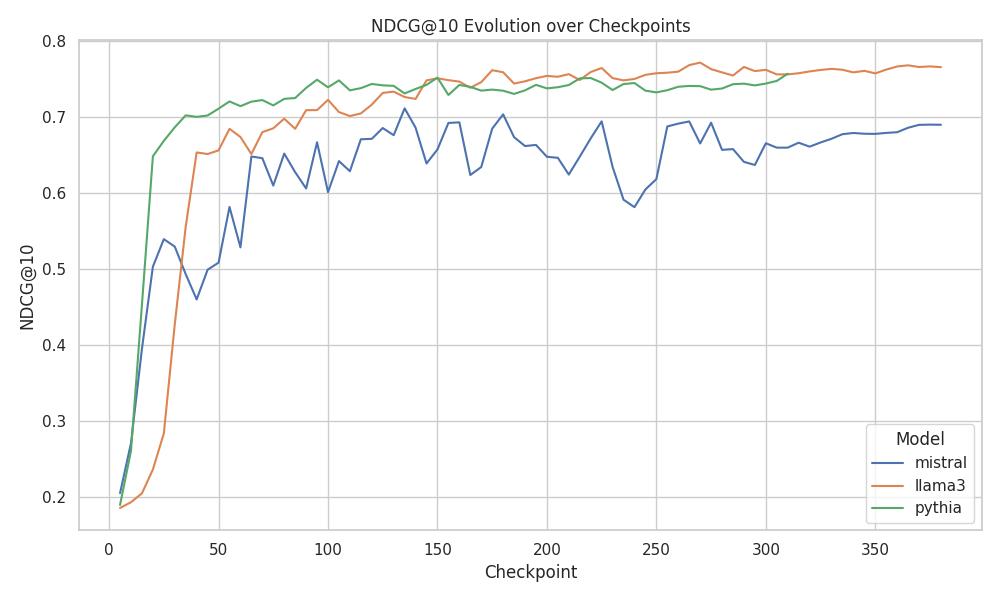}
    \caption{\small NDCG@10 evolution across LoRA fine-tuning checkpoints for LLaMa3, Mistral, and Pythia models (rank = 8). Relevance improves rapidly within the first ~50 steps and stabilizes after ~300 steps.}
    \label{fig:ndcg_evolution}
\end{figure}

\begin{figure}[h]
    \centering
    \includegraphics[width=0.5\linewidth]{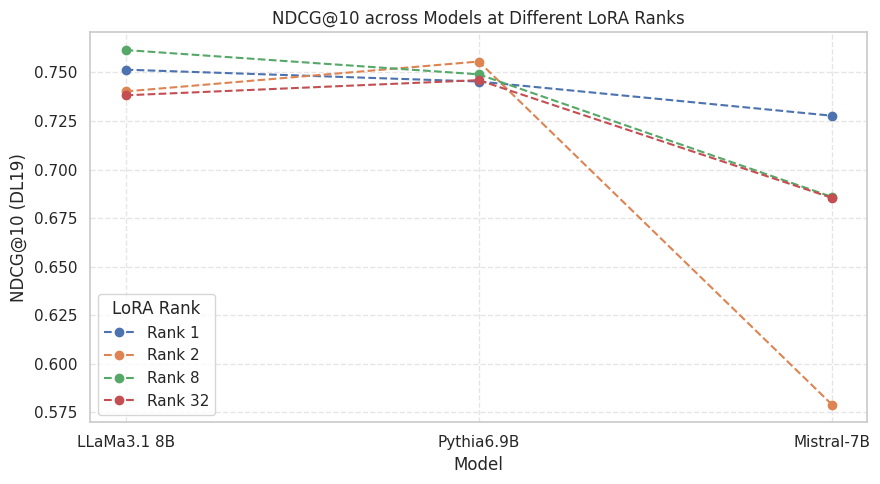}
    \caption{\small NDCG@10 scores for rerankers fine-tuned with different LoRA ranks (1, 2, 8, 32). Scores are closely clustered, indicating limited sensitivity to rank choice for LLaMA and Pythia models.}
    \label{fig:ndcg_ranks}
\end{figure}

\begin{figure}[h]
    \centering
    \includegraphics[width=0.5\linewidth]{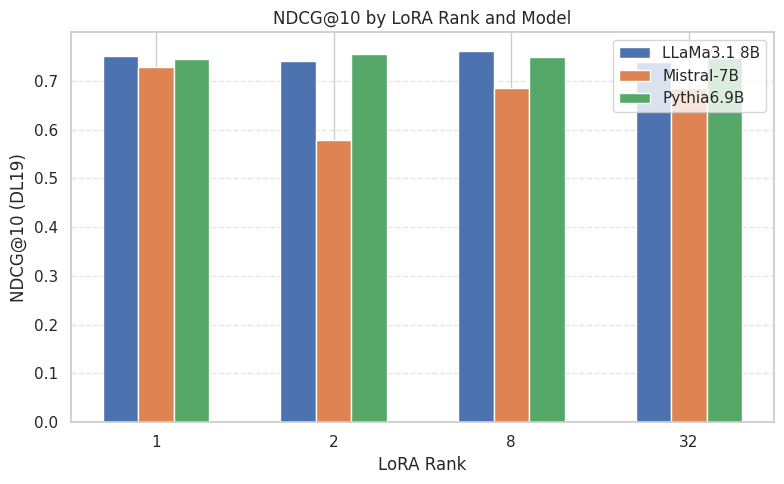}
    \caption{\small NDCG@10 scores for rerankers fine-tuned with different LoRA ranks (1, 2, 8, 32). Scores are closely clustered, indicating limited sensitivity to rank choice.}
    \label{fig:ndcg_ranks_bar}
\end{figure}

\begin{figure}[h]
    \centering
    \includegraphics[width=0.5\linewidth]{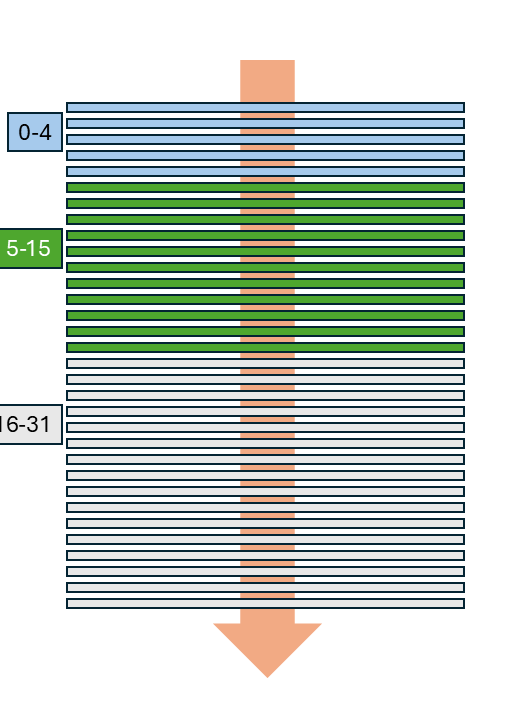}
    \caption{\small Effect of LoRA updates applied to different layer ranges and components (MLP-only) in LLaMA3, Mistral, and Pythia models. Updating only layers 5–15 retains 88–92\% of full performance, while updating only the Up+Gate projections recovers ~96\%.}
    \label{fig:32_layers}
\end{figure}

\vspace{-1.5mm}
We examine how relevance modeling evolves in LoRA fine-tuning, the influence of LoRA rank, the distinct roles of MHA and MLP updates, and layer-wise contributions of LoRA to final predictions. All of these results are from the test set (DL19).
\begin{description}[left=0pt]
    \item[Emerging relevance during LoRA fine-tuning:] We track how reranking performance evolves across fine-tuning checkpoints, as shown in Table~\ref{tab:evo_finetuning}. All models improve steadily over time, with LLaMA3 and Pythia surpassing Mistral by Step $50$ in terms of MRR and NDCG@10. By Step $300$, LLaMA3 slightly edges out Pythia, indicating stronger late-stage performance, although all models converge by then. Please refer to the \textit{figure~\ref{fig:ndcg_evolution}} for how NDCG@10 evolves across fine-tuning checkpoints for each base model fine-tuned with LoRA rank 8.
    \item[Effect of varying LoRA rank on
model adaptation:] We experiment with ranks 1,2,8,32 to fine-tune the LLMs for the passage reranking task. The results are presented in Table \ref{tab:ranks}. We observe that the models achieve comparable ranking performance across different LoRA ranks, suggesting that they adapt similarly irrespective of rank. Notably, a LoRA rank of 1 is already sufficient to encode effective ranking behavior, at least on the MS MARCO dataset. Please refer to \textit{Figure~\ref{fig:ndcg_ranks}} and \textit{Figure~\ref{fig:ndcg_ranks_bar}}  to see how NDCG@10 scores evolve across different LoRA ranks.
    \item[MHA vs MLP LoRa updates:] We perform experiments zeroing out MHA updates and MLP updates of the LoRA fine-tuned model individually, as shown in Table \ref{tab:mha_mlp}. MLP-only and MHA-only updates both significantly improve ranking quality, with MHA giving a stronger boost than MLP. However, combining both leads to the highest MRR and NDCG@10 for all three models, indicating that each component contributes complementary benefits. As we are primarily interested in extracting novel features within LLMs, we here-on only fine-tune on the MLP LoRA components.
    \item[Layer and Projection-Specific Insights in LoRA:] We fine-tune only the MLP layers of LLMs and perform ablation studies by selectively zeroing out specific layers and projections -- i.e., replacing their fine-tuned weights with those from the base model. Our results (Table ~\ref{tab:projection_ablation} and \textit{figure~\ref{fig:32_layers}})  show that layers 5–15 contribute most to relevance prediction because when just those updates are removed, the performance plummets.  We also see that the \emph{up} and \emph{gate} projections are substantially more impactful than the \emph{down} projection in the MLP blocks. Additionally, we find that LoRA updates applied only to the MLP layers can recover up to $98\%$ of the relevance modeling performance achieved by updating both MHA and MLP components. This potentially indicates that ranking specific information is localized to certain layers and projections within the LLM.  

\end{description}

%%%%%%%%%%% Table 1 - Evolution of relevance across fine-tuning

\vspace*{0.5cm}

\begin{table*}
\centering
%\caption{Tracking Relevance score across LoRA fine tuning checkpoints (steps). Fine tuned models are RankLLaMa3 \cite{rankllama3_r8}, RankMistral \cite{rankmistral_r8}  and RankPythia \cite{rankpythia_r8} }
\caption{Tracking Relevance score across LoRA fine tuning checkpoints (steps) for Llama3, Mistral and Pythia.}
\vspace{-3mm}
\begin{tabular}{|c||ccc|ccc|ccc|}
\hline
\multirow{2}{*}{Step} & \multicolumn{3}{c|}{LLaMA3 } & \multicolumn{3}{c|}{Mistral} & \multicolumn{3}{c|}{Pythia} \\
 & MAP & MRR & NDCG@10 & MAP & MRR & NDCG@10 & MAP & MRR & NDCG@10 \\ \hline \hline
5   & 0.1887 & 0.4220 & 0.1855 & 0.1969 & 0.4147 & 0.2050 & 0.1955 & 0.4339 & 0.1895 \\
10  & 0.1932 & 0.4485 & 0.1931 & 0.2350 & 0.5224 & 0.2708 & 0.2296 & 0.4947 & 0.2606 \\
%20  & 0.2103 & 0.4838 & 0.2361 & 0.3520 & 0.6647 & 0.5030 & 0.4325 & 0.9474 & 0.6483 \\
%30  & 0.3227 & 0.7389 & 0.4271 & 0.3903 & 0.7126 & 0.5294 & 0.4626 & 0.9615 & 0.6864 \\
%50  & 0.4553 & 0.8845 & 0.6560 & 0.3472 & 0.7684 & 0.5083 & 0.4757 & 0.9438 & 0.7109 \\
100 & 0.4861 & 0.9380 & 0.7225 & 0.4094 & 0.7657 & 0.6010 & 0.4957 & 0.9884 & 0.7390 \\
%200 & 0.5057 & 0.9826 & 0.7540 & 0.4618 & 0.8672 & 0.6476 & 0.5081 & 0.9523 & 0.7376 \\
300 & 0.5141 & 0.9709 & 0.7620 & 0.4652 & 0.8948 & 0.6654 & 0.5104 & 0.9814 & 0.7439 \\
\hline
\end{tabular}
\label{tab:evo_finetuning}
\end{table*}

\vspace*{0.5cm}

\vspace*{0.5cm}

%%%%%%Table 2 - Experiments with different LoRA ranks
\begin{table}[ht]
    \centering
    \caption{NDCG@10  on DL-19 for LLMs fine tuned with different LoRA Ranks, along with Checkpoint at which they converge.}
    \vspace{-3mm}
    \label{tab:ndcg_transposed}
    \begin{tabular}{|c||c|c|c|}
        \hline
        \multirow{2}{*}{\textbf{Rank}} & \multicolumn{3}{c|}{\textbf{Base Model}} \\ \cline{2-4}
        & \textbf{LLaMa3.1 8B} & \textbf{Mistral-7B} & \textbf{Pythia6.9b} \\
        \hline \hline
        1  & 0.7514 (150) & 0.7277 (200)     & 0.7453 (75)           \\ \hline
        2  & 0.7402 (100) & 0.5790 (400)     & 0.7556 (100)           \\ \hline
        8  & 0.7615 (175) & 0.6859 (140) & 0.7490 (95)  \\ \hline
        32 & 0.7382 (175) & 0.6854 (125) & 0.7460 (75)  \\ \hline
    \end{tabular}
    \label{tab:ranks}
\end{table}

%%%%%%%%%%%%Table 3 - MHA/MLP ablation
\begin{table}[ht]
    \centering
    \vspace{-3mm}
    \caption{Evaluation after zeroing out MLP or MHA components across all layers in a reranker model finetuned with LoRA Rank 8.}
    \vspace{-3mm}
    \label{tab:model_edits_transposed}
    \begin{tabular}{|c||cc|cc|cc|}
        \hline
        \multirow{2}{*}{\textbf{Setting}} & \multicolumn{2}{c|}{\textbf{LLaMa3}} & \multicolumn{2}{c|}{\textbf{Mistral}} & \multicolumn{2}{c|}{\textbf{Pythia}} \\ \cline{2-7}
        & MRR & NDCG@10 & MRR & NDCG@10 & MRR & NDCG@10 \\
        \hline \hline
        No LoRA updates & 0.422 & 0.1855 & 0.2978 & 0.1450 & 0.4339 & 0.1895 \\ \hline
        MLP updated only & 0.6971 & 0.4341 & 0.5899 & 0.4390 & 0.9048 & 0.6262 \\ \hline
        MHA updated only & 0.8508 & 0.5987 & 0.7101 & 0.5164 & 0.9593 & 0.7009 \\ \hline
        Both MLP \& MHA updated & 0.9543 & 0.7655 & 0.9050 & 0.6891 & 0.9709 & 0.7569 \\ \hline        
    \end{tabular}
    \label{tab:mha_mlp}
\end{table}

%%%%%%%%Table 4 ---- Projection Ablation
\begin{table}[h]
    \centering
    \vspace{-3mm}
%    \caption{Evaluation with various layer/projection ablations with MLP-only LoRA models \cite{rankmistral_r8_mlp_only} \cite{rankpythia_r8_mlp_only} \cite{rankllama3_r8_lora_mlp} on DL19 (NDCG@10). }
    \caption{Evaluation with various layer/projection ablations with MLP-only LoRA models on DL19 (NDCG@10). }

    \vspace{-3mm}
    \label{tab:ndcg_results_mlp_only_lora}
    \begin{tabular}{|l||c|c|c|||l||c|c|c|}
        \hline 
        \textbf{LoRA Updates} & \textbf{LLaMA} & \textbf{Mistral} & \textbf{Pythia} 
        & \textbf{LoRA Updates} & \textbf{LLaMA} & \textbf{Mistral} & \textbf{Pythia} \\
        \hline \hline
        All Layers
        & 0.7497 & 0.7318 & 0.7570 
        & Layers \textcolor{darkpastelblue}{0 to 4} and \textcolor{gray}{16 to 31} & 0.2900 & 0.2494 & 0.4016 \\
        
        \textcolor{darkspringgreen}{Layers 5 to 15} & 0.6678 & 0.5559 & 0.5895 
        & All Layers Up \& gate\_proj  & 0.7373 & 0.6947 & 0.7262 \\
        
        \textcolor{darkspringgreen}{Layers 5 }\textcolor{gray}{to 31} & 0.7184 & 0.6831 & 0.6192 
        & \textcolor{darkpastelblue}{Layers }\textcolor{darkspringgreen}{0 to 15} & 0.7065 & 0.5640 & 0.7463 \\
        
        % \multicolumn{4}{|l||}{Layers 0 to 4 and 16 to 31, down\_proj} & 0.5025 & 0.4844 & 0.3898 & \\ All modules in 
        \hline
    \end{tabular}
    \label{tab:projection_ablation}
\end{table}

\vspace*{0.5cm}

\vspace{-3mm}
\section{Conclusion}
\vspace{-1.5mm}
Our experiments provided a detailed glance into where and how relevance signals emerge in LoRA-based fine-tuning for passage reranking. We found that even a LoRA rank of $1$ can capture effective ranking behavior, that MHA and MLP updates contribute complementary gains (with MHA offering a stronger boost), that mid-range MLP layers (especially up and gate projections) are most critical for encoding relevance and that MLP only LoRA updates can recover upto $98\%$ of relevance information. These insights explain the role of of low-rank adaptations in IR, and they pave the way for further mechanistic investigations into how LLMs learn and store retrieval-specific knowledge.

\bibliographystyle{ACM-Reference-Format}

\bibliography{sample-base}

%%% -*-BibTeX-*-
%%% Do NOT edit. File created by BibTeX with style
%%% ACM-Reference-Format-Journals [18-Jan-2012].

\begin{thebibliography}{7}

%%% ====================================================================
%%% NOTE TO THE USER: you can override these defaults by providing
%%% customized versions of any of these macros before the \bibliography
%%% command.  Each of them MUST provide its own final punctuation,
%%% except for \shownote{} and \showURL{}.  The latter two
%%% do not use final punctuation, in order to avoid confusing it with
%%% the Web address.
%%%
%%% To suppress output of a particular field, define its macro to expand
%%% to an empty string, or better, \unskip, like this:
%%%
%%% \newcommand{\showURL}[1]{\unskip}   % LaTeX syntax
%%%
%%% \def \showURL #1{\unskip}           % plain TeX syntax
%%%
%%% ====================================================================

\ifx \showCODEN    \undefined \def \showCODEN     #1{\unskip}     \fi
\ifx \showISBNx    \undefined \def \showISBNx     #1{\unskip}     \fi
\ifx \showISBNxiii \undefined \def \showISBNxiii  #1{\unskip}     \fi
\ifx \showISSN     \undefined \def \showISSN      #1{\unskip}     \fi
\ifx \showLCCN     \undefined \def \showLCCN      #1{\unskip}     \fi
\ifx \shownote     \undefined \def \shownote      #1{#1}          \fi
\ifx \showarticletitle \undefined \def \showarticletitle #1{#1}   \fi
\ifx \showURL      \undefined \def \showURL       {\relax}        \fi
% The following commands are used for tagged output and should be
% invisible to TeX
\providecommand\bibfield[2]{#2}
\providecommand\bibinfo[2]{#2}
\providecommand\natexlab[1]{#1}
\providecommand\showeprint[2][]{arXiv:#2}

\bibitem[Chowdhury and Allan(2024)]%
        {chowdhury2024probing}
\bibfield{author}{\bibinfo{person}{Tanya Chowdhury} {and} \bibinfo{person}{James Allan}.} \bibinfo{year}{2024}\natexlab{}.
\newblock \showarticletitle{Probing Ranking LLMs: Mechanistic Interpretability in Information Retrieval}.
\newblock \bibinfo{journal}{\emph{arXiv preprint arXiv:2410.18527}} (\bibinfo{year}{2024}).
\newblock


\bibitem[Craswell et~al\mbox{.}(2020)]%
        {craswell2020overview}
\bibfield{author}{\bibinfo{person}{Nick Craswell}, \bibinfo{person}{Bhaskar Mitra}, \bibinfo{person}{Emine Yilmaz}, \bibinfo{person}{Daniel Campos}, {and} \bibinfo{person}{Ellen~M Voorhees}.} \bibinfo{year}{2020}\natexlab{}.
\newblock \showarticletitle{Overview of the TREC 2019 deep learning track}.
\newblock \bibinfo{journal}{\emph{arXiv preprint arXiv:2003.07820}} (\bibinfo{year}{2020}).
\newblock


\bibitem[Hu et~al\mbox{.}(2022)]%
        {hu2022lora}
\bibfield{author}{\bibinfo{person}{Edward~J Hu}, \bibinfo{person}{Yelong Shen}, \bibinfo{person}{Phillip Wallis}, \bibinfo{person}{Zeyuan Allen-Zhu}, \bibinfo{person}{Yuanzhi Li}, \bibinfo{person}{Shean Wang}, \bibinfo{person}{Lu Wang}, \bibinfo{person}{Weizhu Chen}, {et~al\mbox{.}}} \bibinfo{year}{2022}\natexlab{}.
\newblock \showarticletitle{Lora: Low-rank adaptation of large language models.}
\newblock \bibinfo{journal}{\emph{ICLR}} \bibinfo{volume}{1}, \bibinfo{number}{2} (\bibinfo{year}{2022}), \bibinfo{pages}{3}.
\newblock


\bibitem[Ma et~al\mbox{.}(2023)]%
        {ma2023fine}
\bibfield{author}{\bibinfo{person}{Xueguang Ma}, \bibinfo{person}{Liang Wang}, \bibinfo{person}{Nan Yang}, \bibinfo{person}{Furu Wei}, {and} \bibinfo{person}{Jimmy Lin}.} \bibinfo{year}{2023}\natexlab{}.
\newblock \bibinfo{title}{Fine-Tuning LLaMA for Multi-Stage Text Retrieval. CoRR abs/2310.08319 (2023)}.
\newblock


\bibitem[Ma et~al\mbox{.}(2024)]%
        {ma2024fine}
\bibfield{author}{\bibinfo{person}{Xueguang Ma}, \bibinfo{person}{Liang Wang}, \bibinfo{person}{Nan Yang}, \bibinfo{person}{Furu Wei}, {and} \bibinfo{person}{Jimmy Lin}.} \bibinfo{year}{2024}\natexlab{}.
\newblock \showarticletitle{Fine-tuning llama for multi-stage text retrieval}. In \bibinfo{booktitle}{\emph{Proceedings of the 47th International ACM SIGIR Conference on Research and Development in Information Retrieval}}. \bibinfo{pages}{2421--2425}.
\newblock


\bibitem[Nguyen et~al\mbox{.}(2016)]%
        {nguyen2016ms}
\bibfield{author}{\bibinfo{person}{Tri Nguyen}, \bibinfo{person}{Mir Rosenberg}, \bibinfo{person}{Xia Song}, \bibinfo{person}{Jianfeng Gao}, \bibinfo{person}{Saurabh Tiwary}, \bibinfo{person}{Rangan Majumder}, {and} \bibinfo{person}{Li Deng}.} \bibinfo{year}{2016}\natexlab{}.
\newblock \showarticletitle{Ms marco: A human-generated machine reading comprehension dataset}.
\newblock  (\bibinfo{year}{2016}).
\newblock


\bibitem[Shuttleworth et~al\mbox{.}(2024)]%
        {shuttleworth2024lora}
\bibfield{author}{\bibinfo{person}{Reece Shuttleworth}, \bibinfo{person}{Jacob Andreas}, \bibinfo{person}{Antonio Torralba}, {and} \bibinfo{person}{Pratyusha Sharma}.} \bibinfo{year}{2024}\natexlab{}.
\newblock \showarticletitle{Lora vs full fine-tuning: An illusion of equivalence}.
\newblock \bibinfo{journal}{\emph{arXiv preprint arXiv:2410.21228}} (\bibinfo{year}{2024}).
\newblock


\end{thebibliography}

\appendix

\end{document}